# Stable Nonlinear and IQ Imbalance RF Fingerprint for Wireless OFDM Devices

Honglin Yuan[1*], Jiangzhou Wang[2], Chen Xu[1], Juping Gu[1], Qiang Sun[1], Yan Yan[1], Zhihua Bao[1]
( 1. Nantong University, Jiangsu Nantong 226019, China; 2. University of Kent, Canterbury, Kent, CT2 7NT, UK )

***Abstract*—An estimation method of Radio Frequency fingerprint (RFF) based on the physical hardware properties of the nonlinearity and in-phase and quadrature (IQ) imbalance of the transmitter is proposed for the authentication of wireless orthogonal frequency division multiplexing (OFDM) devices. Firstly, the parameters of the nonlinearity of the transmitter and finite impulse response (FIR) of the wireless multipath channel are estimated with a Hammerstein system parameter separation technique. Secondly, the best IQ imbalance parameter combination estimation is obtained with a searching algorithm and the applied conjugate anti-symmetric pilots. Thirdly, the estimations of the nonlinear coefficients and IQ imbalance parameter combination are considered as a novel RFF, the features are extracted from the novel RFF, and a *k*-Nearest Neighbor (*k*-NN) classifier is used to classify the communication devices with the features. It is demonstrated with the numerical experiments of five transmitters that the novel RFF eliminates the adverse effect of the wireless channel and is therefore stable. The proposed RF fingerprinting method is helpful for the high-strength authentication of the OFDM communication devices with subtle differences from the same model and same series.**

## I. INTRODUCTION

AS the rapid development of the techniques of the fifth generation mobile communication and internet of thing etc., the physical-layer information security of communication networks has inevitably become a hot topic. RF Fingerprinting is one of the methods for the enhancement of the physical-layer information security, which relies on the hardware characteristics of the communication devices, RF fingerprint (RFF), rather than the digital information such as cryptographic key etc. to authenticate the real identities of the wireless devices.[1-2]

Although the digital information of the training preamble or ideal symbol constellations of a communication frame is definite, the corresponding analog signals transmitted from different wireless devices are different as their transmitter hardware is unique, even though the devices are from the same model and same series. Therefore, the received signals have widely been used to develop the signal profile related RFFs for the authentication of the wireless devices [3-5], where [3] was the first published paper using the preamble envelope to enhance the security of IEEE 802.11b wireless networks. However, the stability of the signal profile related RFFs is easily damaged by the time-varying wireless multipath fading channel.

On the other hand, the hardware parameters of the transmitter of the wireless device have been estimated as hardware related RFFs which are stable. [6] extracted the RFF based on the imbalance between the in-phase and quadrature (IQ) components of the transmitter. In [7], the amplifier was modeled as a Taylor series, a nonlinear RFF was proposed with the derived carrier component and harmonic component expression of the transmitted signal, which was independent to the base band signal. Classification experiments with 4 FM emitters obtained the successful results. In [8], the wireless transmitter was modeled in a power series expansion format. With the preamble as the input and the corresponding received signal as the output, the coefficients of the nonlinear model was estimated with an iterative algorithm. The estimated coefficients was considered as a nonlinear RFF to authenticate the radio devices. The simulation experiments showed good estimation precision and high classification rates. The previous work [9-10] proposed a nonlinear RFF authentication method which separated the influence of the wireless channel. The previous work [11] proposed an IQ imbalance and nonlinear RFF for OFDM communication based on B-Spline Neural Network, which applied the conjugate symmetric and anti-symmetric pilots. The above hardware related RFF have stable performance but are limited by some conditions.

In this paper, we propose a novel OFDM RF Fingerprinting method which relies on the hardware property of the IQ imbalance and nonlinearity of the transmitter together. The demand of pilot for this novel method is less than that for [11], where the two RFF have all eliminated the adverse effect of the wireless multiple path channel and are feasible to differentiate the transmitters with minor differences.

The system model is given in Section II. The Hammerstein system parameter separation technique based on the previous work [9-10], the searching of the best IQ imbalance parameter combination estimation and novel RF Fingerprinting method are given in Section III. In Section IV, numerical classification experiments of five transmitters with minor parameter differences are given, which demonstrates the feasibility of the proposed method. Section V is the conclusion.

## II. SYSTEM MODEL

The low-pass equivalent model of the RF Fingerprinting system for OFDM communication is shown in Fig. 1, which consists of the OFDM signal generator, digital to analog converter (DAC), IQ modulator, power amplifier (PA), wireless multipath channel and additive white Gaussian noise (AWGN) etc. And the nonlinear part of the transmitter and the linear wireless channel can also be modeled as a Hammerstein system which is also shown in Fig. 1.

* Corresponding author, Email: yuan.hl@ntu.edu.cn, ORCID: 0000-0002-3320-9412.

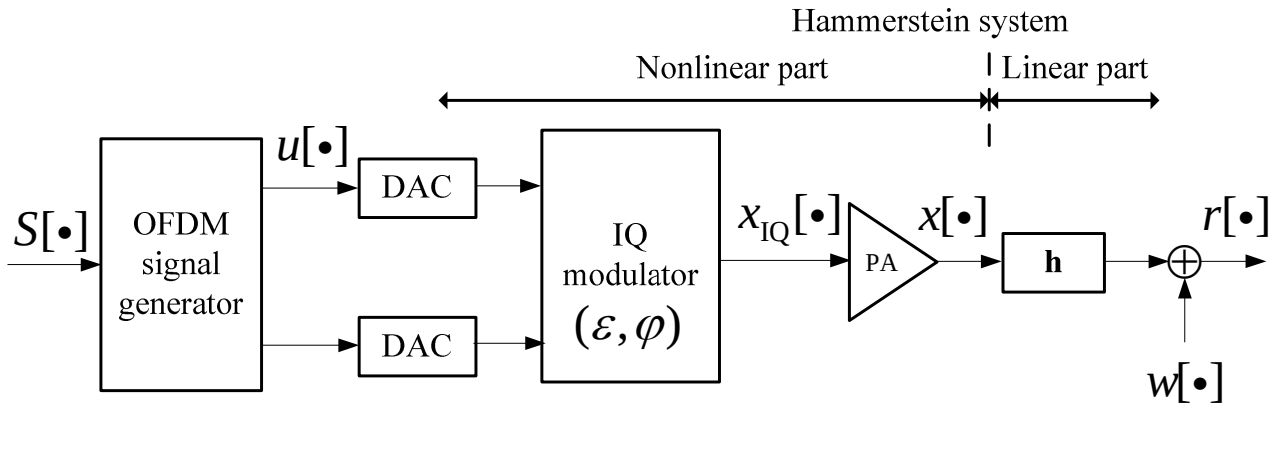

Fig. 1 The low-pass equivalent model of RF Fingerprinting system for OFDM communication

Suppose the known training frequency-domain (FD) symbol vector of an OFDM pilot symbol is denoted as $\mathbf{s}=[S[0],\cdots,S[N-1]]^T$, where *N* is the number of the FD symbols. The standard discrete Fourier transform (DFT) matrix is denoted as **W**. The time-domain (TD) vector corresponding to $\mathbf{s}$ is then described as

$$\begin{aligned}\mathbf{u}&=\frac{1}{N}\mathbf{W}^H\mathbf{s}\\&=[u[0],\cdots,u[N-1]]^T\end{aligned} \tag{1}$$

Suppose the length of the cyclic prefix (CP) of the OFDM symbol is $L_{\mathrm{CP}}$, the TD OFDM symbol vector is then extended as

$$\tilde{\mathbf{u}}=[u[-L_{\mathrm{CP}}],\cdots,u[-1],\mathbf{u}^T]^T \tag{2}$$

where $u[-n]=u[N-n]$, $1\le n\le L_{\mathrm{CP}}$.

The $\varepsilon$ and $\varphi$ in IQ modulator of Fig. 1 denote the amplitude imbalance and phase deviation of the IQ modulator, respectively. Suppose $G_1=(1+\varepsilon e^{j\varphi})/2$ and $G_2=(1-\varepsilon e^{j\varphi})/2$. After $\tilde{\mathbf{u}}$ goes through the IQ modulator, it becomes

$$\begin{aligned}\tilde{\mathbf{x}}_{\mathrm{IQ}}&=G_1\tilde{\mathbf{u}}+G_2\tilde{\mathbf{u}}^*\\&=[x_{\mathrm{IQ}}[-L_{\mathrm{CP}}],\cdots,x_{\mathrm{IQ}}[-1],\mathbf{x}_{\mathrm{IQ}}{}^T]^T\end{aligned} \tag{3}$$

where $\mathbf{x}_{\mathrm{IQ}}=[x_{\mathrm{IQ}}[0],\cdots,x_{\mathrm{IQ}}[N-1]]^T$ and $x_{\mathrm{IQ}}[-n]=x_{\mathrm{IQ}}[N-n]$, $1\le n\le L_{\mathrm{CP}}$.

When $\tilde{\mathbf{x}}_{\mathrm{IQ}}$ passes the PA, it becomes

$$\begin{aligned}\tilde{\mathbf{x}}_{\mathrm{PA}}&=\psi(\tilde{\mathbf{x}}_{\mathrm{IQ}})\\&=[\psi(x_{\mathrm{IQ}}[-L_{\mathrm{CP}}]),\cdots,\psi(x_{\mathrm{IQ}}[-1]),\psi(\mathbf{x}_{\mathrm{IQ}})^T]^T\\&=[x[-L_{\mathrm{CP}}],\cdots,x[-1],\mathbf{x}^T]^T\end{aligned} \tag{4}$$

where the element

$$\begin{aligned}x[n]&=\psi(x_{\mathrm{IQ}}[n])\\&=A(|x_{\mathrm{IQ}}[n]|)e^{j\cdot(\angle x_{\mathrm{IQ}}[n]+\theta(|x_{\mathrm{IQ}}[n]|))}\end{aligned} \tag{5}$$

and the vector

$$\begin{aligned}\mathbf{x}&=\psi(\mathbf{x}_{\mathrm{IQ}})\\&=[x[0],\cdots,x[N-1]]^T\\&-L_{\mathrm{CP}}\le n\le N-1\end{aligned} \tag{6}$$

The $\psi(\bullet)$ in (4) ~ (6) denotes the memoryless nonlinearity of the PA in the transmitter, $A(\bullet)$ and $\theta(\bullet)$ in (5) are the amplitude and phase response of $\psi(\bullet)$, respectively, and $|\bullet|$ and $\angle\bullet$ are the amplitude and angle of $\bullet$, respectively. Suppose the amplitude of the input of the PA is *r* and

$$\begin{aligned}A(r)&=\frac{g_\alpha r}{(1+(\frac{g_\alpha r}{A_{sat}})^{2\beta_\alpha})^{\frac{1}{2\beta_\alpha}}}\\\theta(r)&=\frac{\alpha_\theta r^{q_1}}{1+(\frac{r}{\beta_\theta})^{q_2}}\end{aligned} \tag{7}$$

where $g_\alpha$, $\beta_\alpha$, $A_{sat}$ are the parameters for the specific amplitude characteristics, and $\alpha_\theta$, $\beta_\theta$, $q_1$ and $q_2$ are for the phase response of the PA[12].

Suppose the vector of the finite impulse response (FIR) of the multipath channel is

$$\mathbf{h}=[h_0,\cdots h_L]^T \tag{8}$$

where $L+1<L_{\mathrm{CP}}$. Suppose $h_0=1$. Then, the received signal vector cut the tail of linear convolution of the multipath channel is denoted as $\tilde{\mathbf{r}}=[r[-L_{\mathrm{CP}}],\cdots,r[-1],\mathbf{r}^T]^T$ where $\mathbf{r}=[r[0],\cdots,r[N-1]]^T$. The element of **r** is denotes as

$$r[n]=\sum_{k=0}^{L-1}h_k\psi(x_{\mathrm{IQ}}[n-k])+w[n] \tag{9}$$

where *w*[*n*] is the AWGN, $-L_{\mathrm{CP}}\le n\le N-1$. And the received signal removing the CP is

$$r[n]=\sum_{k=0}^{N-1}\psi(x_{IQ}[k])h_{((n-k))_N}R_N(n)+w[n] \tag{10}$$

where $0\le n\le N-1$, $h_{((n-k))_N}$ is the extension with period *N* of the reverse of $h_k$ and the multiplication of $R_N(n)$ is to obtain the main values of $h_{((n-k))_N}$. Then,

$$\begin{aligned}\mathbf{r}&=\mathbf{H}\psi(\mathbf{x}_{\mathrm{IQ}})+\mathbf{w}\\&=[r[0],\cdots r[N-1]]^T\end{aligned} \tag{11}$$

where $\mathbf{w}=[w[0],\cdots,w[N-1]]^T$ and

$$\mathbf{H} = \begin{bmatrix} h_0 & 0 & \cdots & 0 & h_L & \cdots & h_2 & h_1 \\ h_1 & h_0 & \cdots & 0 & 0 & \cdots & h_3 & h_2 \\ \vdots & \vdots & & \vdots & \vdots & & \vdots & \vdots \\ 0 & 0 & \cdots & h_L & h_{L-1} & \cdots & h_1 & h_0 \end{bmatrix} \quad (12)$$

is the circulation matrix of the FIR of the multipath channel.

## III. THE NOVEL RF FINGERPRINTING METHOD

### *A. The Hammerstein system parameter separation technique*

Th received signal of the OFDM symbol is used to estimate the nonlinear parameter of the transmitter and FIR of multipath channel with the Hammerstein system parameter separation technique which is described as follows.

As shown in Fig. 1, after the TD OFDM signal $u[n]$, $-L_{\mathrm{CP}} \le n \le N-1$, passes through the IQ modulator and PA, it can also be described as

$$x[n] = \sum_{p=1}^{(P+1)/2} b_{2p-1}\phi_p(u[n]) \quad (13)$$

where $P$ is an odd number, $\phi_p(\bullet) = \bullet|\bullet|^{2(p-1)}$ is the conventional polynomial basis function, and $b_i, i = 1,3,\cdots,P$ are the coefficients. Then, the received signal can be written as

$$r[n] = \sum_{l=0}^{L} \phi_{\mathbf{p},u}{}^T(n-l)h_l\mathbf{b} + w[n] \quad (14)$$

where $\phi_{\mathbf{p},u}(\bullet) = [\phi_1(u[\bullet]),...,\phi_{\frac{P+1}{2}}(u[\bullet])]^T$ is the vector composed of the conventional polynomial basis functions of the signal $u[\bullet]$, and $\mathbf{b} = [b_1, b_3, \cdots, b_P]^T$ is the vector composed of the coefficients of the nonlinear conventional polynomial model. It can be seen with (14) that $r[n]$ without AWGN is the linear convolution of vector $\phi_{\mathbf{p},u}(\bullet)$ and $h_\bullet\mathbf{b}$. Suppose $z^{-1}$ denotes the unit delay, the linear vector model of the nonlinear Hammerstein system in Fig. 1 can be depicted as Fig. 2.

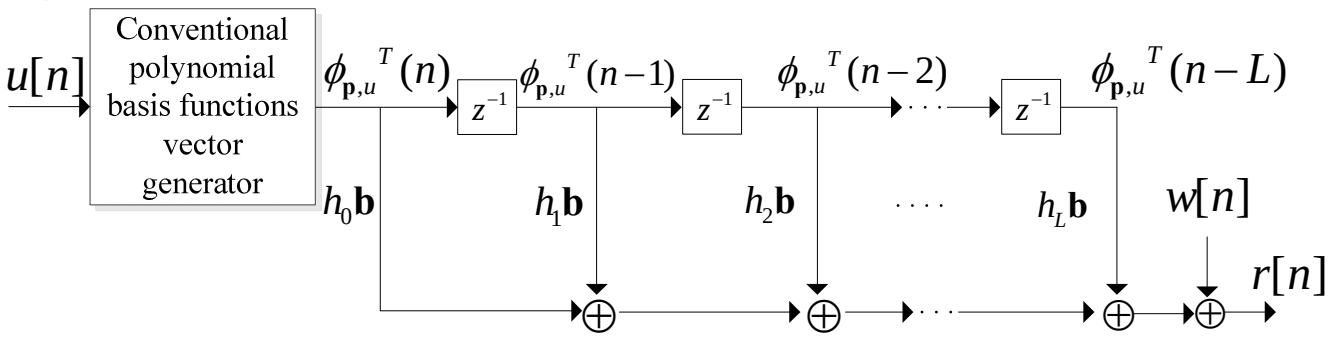

Fig. 2 The linear vector model of the nonlinear Hammerstein system

As in Fig. 2, after the conventional polynomial basis functions vector is generated with the TD OFDM signal, the nonlinear Hammerstein system is converted to a general linear filter with the vectors of $\phi_{\mathbf{p},u}(\bullet)$ and $h_\bullet\mathbf{b}$.

Similar to [13], define the $y$-delayed vector of $\tilde{\mathbf{u}}$ is $\tilde{\mathbf{u}}_y = [\mathbf{0}_{1\times y}, u[-L_{\mathrm{CP}}],...,u[N-1-y]]^T$, $0 \le y \le L$. Then, $\tilde{\mathbf{u}}_0 = \tilde{\mathbf{u}}$. Define the polynomial basis function of $\tilde{\mathbf{u}}_y$ as $\phi_p(\tilde{\mathbf{u}}_y) = [\mathbf{0}_{1\times y}, \phi_p(u[-L_{\mathrm{CP}}]),\cdots,\phi_p(u[N-1-y])]^T$, $1 \le p \le \frac{P+1}{2}$. Define the matrix $\mathbf{\Phi}_y = [\phi_1(\tilde{\mathbf{u}}_y),\cdots,\phi_{\frac{P+1}{2}}(\tilde{\mathbf{u}}_y)]$ and the Kronecker product of the vector $\mathbf{b}$ and $\mathbf{h}$ as $\mathbf{h_b} = [h_0\mathbf{b}^T, h_1\mathbf{b}^T,...,h_L\mathbf{b}^T]^T$. The received signal vector can then be written as

$$\begin{aligned}\tilde{\mathbf{r}} &= \mathbf{\Phi}\mathbf{h_b} + \mathbf{w} \\ &= \mathbf{\Phi}(\mathbf{I}\otimes\mathbf{b})\mathbf{h} + \mathbf{w}\end{aligned} \quad (15)$$

where $\mathbf{\Phi} = [\mathbf{\Phi}_0,\cdots,\mathbf{\Phi}_L]$ and $\otimes$ denotes the operation of Kronecker product.

With (15), it can be seen that the least square (LS) estimation of $\mathbf{h_b}$ is $\hat{\mathbf{h}}_\mathbf{b} = (\mathbf{\Phi}^H\mathbf{\Phi})^{-1}\mathbf{\Phi}^H\tilde{\mathbf{r}}$. Let us obtain the elements of $\hat{\mathbf{h}}_\mathbf{b}$ from number 1 to $(P+1)/2$, which is denoted as $\hat{h}_0\mathbf{b} = \hat{\mathbf{h}}_\mathbf{b}(1:(P+1)/2)$, as $h_0$ is supposed as 1, it can be obtained that the estimation of $\mathbf{b}$ is $\hat{\mathbf{b}} = \hat{h}_0\mathbf{b}$. Calculate $\widehat{\mathbf{\Phi}}_\mathbf{b} = \mathbf{\Phi}(\mathbf{I}\otimes\hat{\mathbf{b}})$ with $\hat{\mathbf{b}}$, where $\mathbf{I}$ is a unit matrix of size $(L+1)\times(L+1)$. Then, $\hat{\mathbf{h}} = (\widehat{\mathbf{\Phi}}_\mathbf{b}{}^H\widehat{\mathbf{\Phi}}_\mathbf{b})^{-1}\widehat{\mathbf{\Phi}}_\mathbf{b}{}^H\tilde{\mathbf{r}}$ is the LS estimation of $\mathbf{h}$.

$\hat{\mathbf{b}}$ and $\hat{\mathbf{h}}$ are the estimations of the nonlinear model parameters and FIR of multipath channel of the equivalent Hammerstein system in Fig. 1 with the received signal of the OFDM symbol and corresponding known training symbols.

### *B. The searching of the best IQ imbalance parameter combination estimation*

The IQ imbalance parameter combination, $\varepsilon e^{j\varphi}$, of the IQ modulator in the transmitter is used for the proposed RFF in this paper. A FD conjugate anti-symmetry pilot is used to compensate the IQ mismatch in CO-OFDM system in [14], which is used here to estimate $\varepsilon e^{j\varphi}$. Suppose that the subset $\mathbf{s}_{\mathrm{B}} = -\overline{\mathbf{s}_{\mathrm{B}}}^*$ is the conjugate anti-symmetry part in the OFDM pilot symbol vector $\mathbf{s}$, which is arranged in the odd or even number of the pilot symbol. We also use the technique of the best linear approximation of the nonlinear PA in [15] to search the best estimation of $\varepsilon e^{j\varphi}$, where the received signal vector removed CP $\mathbf{r}$ is used.

The nonlinear PA is approximated as a linear one, that is denoted as

$$\begin{aligned}\mathbf{x} &= \psi(\mathbf{x}_{\mathrm{IQ}}) \\ &\approx K\mathbf{x}_{\mathrm{IQ}}\end{aligned} \quad (16)$$

where $K$ is a constant. With (11), the best linear approximation $K$ can then be searched as

$$\widehat{K} = \underset{K\in\Im}{argmin}\{\| \mathbf{r} - \mathbf{H}K\mathbf{x}_{\mathrm{IQ}} \|^2\}$$
$$= \underset{K_m\in\Im}{argmin}\{\| \mathbf{r} - \mathbf{H}K[\frac{1}{N}G_1\mathbf{W}^H\mathbf{s} + \frac{1}{N}G_2(\mathbf{W}^H\mathbf{s})^*] \|^2\} \tag{17}$$

where $\|\bullet\|$ denotes the euclidean distance of vector $\bullet$, and the set $\Im$ is determined by the priori knowledge of the PA. Then, the DFT of (11) can be written as

$$\mathrm{DFT}\{\mathbf{r}\} \approx \mathbf{WH}(\widehat{K}\mathbf{x}_{\mathrm{IQ}}) + \mathrm{DFT}\{\mathbf{w}\}$$
$$= \frac{\widehat{K}\mathbf{\Lambda}}{2}[(\mathbf{s}+\overline{\mathbf{s}}^*) + \varepsilon e^{j\varphi}(\mathbf{s}-\overline{\mathbf{s}}^*)] + \mathrm{DFT}\{\mathbf{w}\} \tag{18}$$

where $\mathbf{WHW}^H = N\mathbf{\Lambda}$, $\mathbf{\Lambda} = diag\{\mathrm{DFT}\{\mathbf{h}\}\}$, $diag\{\bullet\}$ denotes the diagonal matrix with the elements of $\bullet$ as the diagonal elements, and $\overline{\mathbf{s}} = [S(0), S(N-1), \cdots, S(1)]^T$ is the mirror image of $\mathbf{s}$. For the subset $\mathbf{s}_{\mathrm{B}}$ of $\mathbf{s}$, (18) is

$$\mathrm{DFT}\{\mathbf{r}_{\mathrm{B}}\} \approx \varepsilon e^{j\varphi}\mathbf{\Lambda}_{\mathrm{B}}\mathbf{s}_{\mathrm{B}}\widehat{K} + \mathrm{DFT}\{\mathbf{w}_{\mathrm{B}}\} \tag{19}$$

where $\bullet_{\mathrm{B}}$ is the corresponding part of $\bullet$ with $\mathbf{s}_{\mathrm{B}}$. Then the best estimation of $\varepsilon e^{j\varphi}$ is

$$\widehat{\varepsilon e^{j\varphi}} \approx \frac{1}{\widehat{K}}\mathrm{E}\{\mathrm{DFT}\{\mathbf{r}_{\mathrm{B}}\}./(\mathbf{\Lambda}_{\mathrm{B}}\mathbf{s}_{\mathrm{B}})\} \tag{20}$$

where $./$ denotes the dot divide of the elements of two vectors and $\mathrm{E}\{\bullet\}$ denotes the operation of the mean of the elements of a vector $\bullet$.

It can be seen with (20) that the best estimation of $\varepsilon e^{j\varphi}$ is determined by $\mathbf{\Lambda}_{\mathrm{B}}$, the best linear approximation of the PA $\widehat{K}$, the conjugate anti-symmetry pilot symbol subset $\mathbf{s}_{\mathrm{B}}$ and its received signal $\mathbf{r}_{\mathrm{B}}$ together, and $\mathbf{\Lambda}_{\mathrm{B}}$ is estimated with the DFT of the estimation of the multipath channel $\hat{\mathbf{h}}$.

*C. The novel RFF and feature extraction*

The estimations of the nonlinear coefficients, $\hat{\mathbf{b}}$, and best IQ imbalance parameter combination estimation, $\widehat{\varepsilon e^{j\varphi}}$, are considered as the novel RFF in this paper.

The ahead 3 nonlinear model coefficients of the transmitters are extracted as the features which compose the vector $\hat{\mathbf{b}} = [\hat{b}_1, \hat{b}_3, \hat{b}_5]^T$. The absolute of $\hat{\mathbf{b}}$ is divided by that of $\hat{b}_1$, which becomes $|\underline{\hat{\mathbf{b}}}| = [1, |\hat{b}_3|/|\hat{b}_1|, |\hat{b}_5|/|\hat{b}_1|]^T$. The 2 dimension feature vectors are then obtained with the elements of $|\underline{\hat{\mathbf{b}}}|$ and $\widehat{\varepsilon e^{j\varphi}}$, which is written as $\mathbf{F}_1 = \{|\hat{b}_3|/|\hat{b}_1|, \widehat{\varepsilon e^{j\varphi}}\}$ and $\mathbf{F}_2 = \{|\hat{b}_5|/|\hat{b}_1|, \widehat{\varepsilon e^{j\varphi}}\}$, respectively.

For the performance comparison, the absolute of the received training signal is obtained as the classical envelope RFF, which is denoted as $|\tilde{\mathbf{r}}|$. The resemble coefficients features are further extracted as follows [16]

$$F_{\mathrm{c}} = \frac{<|\tilde{\mathbf{r}}|, \mathbf{c}>}{\|\tilde{\mathbf{r}}\| \bullet \|\mathbf{c}\|}, F_{\mathrm{i}} = \frac{<|\tilde{\mathbf{r}}|, \mathbf{i}>}{\|\tilde{\mathbf{r}}\| \bullet \|\mathbf{i}\|} \tag{21}$$

where **c** and **i** denote the rectangle and triangle vector with the same length of $|\tilde{\mathbf{r}}|$ respectively, and $<\bullet,\bullet>$ denotes the operation of inner product. $F_{\mathrm{c}}$ and $F_{\mathrm{i}}$ are combined as another feature vector $\mathbf{F}_{\mathrm{e}} = \{F_{\mathrm{c}}, F_{\mathrm{i}}\}$ for the classification performance comparison.

*D. The steps for the novel RF Fingerprinting method*

The steps of the proposed RF Fingerprinting method is summarized as follows:

**Step 1**: obtain the received signal of the training frame $\tilde{\mathbf{r}}$, remove the CP from $\tilde{\mathbf{r}}$, the remaining signal is denoted as **r**. Extract the signal of **r** corresponding to the conjugate anti-symmetry pilots $\mathbf{r}_{\mathrm{B}}$;

**Step 2**: using the Hammerstein system parameter separation technique, estimate the nonlinear model coefficients vector of the transmitter $\hat{\mathbf{b}}$ and FIR of the multipath channel $\hat{\mathbf{h}}$ with $\tilde{\mathbf{r}}$ and the corresponding OFDM TD signal of the training pilots $\tilde{\mathbf{u}}$;

**Step 3**: construct $\hat{\mathbf{\Lambda}}_{\mathrm{B}}$ and the channel circulation matrix $\widehat{\mathbf{H}}$ with $\hat{\mathbf{h}}$;

**Step 4**: set $\Im = K_1 : K_M$ and *m*=1;

**Repeat**:

Set $\widehat{K} = K_m$;

Estimate $\widehat{\varepsilon e^{j\varphi}}_m = \frac{1}{\widehat{\mathrm{K}}}\mathrm{E}\{\mathrm{DFT}\{\mathbf{r}_{\mathrm{B}}\}./(\hat{\mathbf{\Lambda}}_{\mathrm{B}}\mathbf{s}_{\mathrm{B}})\}$;

Calculate $\widehat{G}_1 = \frac{1+\widehat{\varepsilon e^{j\varphi}}_m}{2}, \widehat{G}_2 = \frac{1-\widehat{\varepsilon e^{j\varphi}}_m}{2}$;

Calculate the cost function $\|\mathbf{w}\|^2{}_m = \|\mathbf{r} - \widehat{\mathbf{H}} \bullet \widehat{K} \bullet [\frac{1}{N}\widehat{G}_1\mathbf{W}^H\mathbf{s} + \frac{1}{N}\widehat{G}_2(\mathbf{W}^H\mathbf{s})^*]\|^2$;

Set m = m +1;

**Until** m = M+1;

**Step 5**: search the minimum of $\|\mathbf{w}\|^2{}_m$ which is denoted as $\min(\|\mathbf{w}\|^2{}_m) = \|\mathbf{w}\|^2{}_q$, then $\widehat{\varepsilon e^{j\varphi}} = \widehat{\varepsilon e^{j\varphi}}_q$;

**Step 6**: Construct the feature vectors $\mathbf{F}_1 = \{|\hat{b}_3|/|\hat{b}_1|, \widehat{\varepsilon e^{j\varphi}}\}$ and $\mathbf{F}_2 = \{|\hat{b}_5|/|\hat{b}_1|, \widehat{\varepsilon e^{j\varphi}}\}$;

**Step 7**: Construct the feature vector $\mathbf{F}_{\mathrm{e}} = \{F_{\mathrm{c}}, F_{\mathrm{i}}\}$ with the received signal $\tilde{\mathbf{r}}$;

**Step 8**: Using a classification tool, *k*-Nearest Neighbor (*k*-NN) etc., to classify the RF devices with $\mathbf{F}_1$, $\mathbf{F}_2$ and $\mathbf{F}_{\mathrm{e}}$, respectively.

## IV. Numerical Experiments

Numerical classification experiments with five transmitters are done to verify the feasibility of the proposed RF Fingerprinting method. The five transmitters to be classified, wireless multipath channel and AWGN are made based on the system model of Fig. 1. Seven IQ imbalance and PA parameters are set to emulate the transmitters with minor differences from the same model and same series. The TD OFDM training signal containing the conjugate anti-symmetry pilots is generated, where the bit mapping scheme is 16-QAM, the length of FFT is 2048, and $L_{\mathrm{CP}}$ =512. The Rayleigh fading channel is generated as the wireless multipath channel, whose maximum channel delay is 9 samples and the first path is normalized as 1. Although the Rayleigh channel is time variant, the power of each channel keeps constant during a transmission of the training signal.

### *A. The hardware parameter setting of five transmitters*

Suppose the five transmitters to be classified are denoted as Transmitter-*x*, $x = 1,2,\cdots,5$, respectively. The hardware parameters of the nonlinear PA in (7) are defined as

$$\begin{aligned} &\beta_\alpha = 0.81(1+\Delta), A_{sat} = 1.4(1+\Delta), \\ &\beta_\theta = 0.123(1+\Delta), q_1 = 3.8(1+\Delta), \\ &q_2 = 3.7(1+\Delta) \end{aligned} \tag{22}$$

where $\Delta$ is the common parameter for the PA. The parameters for the nonlinear PA and IQ imbalance in modulator are set in Table 1.

TABLE I
The IQ imbalance and PA parameters of the five transmitters where -x denotes Transmitter-x, x=1,2,3,4,5

| Parameter | -1 | -2 | -3 | -4 | -5 |
|---|---|---|---|---|---|
| $\varepsilon$ | 0.99 | 0.96 | 0.97 | 0.90 | 0.93 |
| $\varphi(^\circ)$ | 2.50 | 1.00 | 0.50 | 2.00 | 1.40 |
| $\Delta$ | 0.01 | 0.16 | 0.31 | 0.11 | 0.26 |

When the IQ modulator of the transmitter is balance, $\varepsilon = 1.00$ and $\varphi = 0.00^\circ$. It can be seen with Table 1 that the differences of the hardware of the five transmitters are subtle.

The nonlinear characteristics of the five PAs are shown in Fig. 3 (a) ~ (b), which describe the AM-AM and AM-PM characteristics of the five PAs respectively.

PA-*x*, $x = 1,2,\cdots,5$, in Fig. 3 denotes the PA of Transmitter-*x*, respectively. It can be seen with Fig. 3 that there are some differences existing between each pair of two PAs of the five transmitters.

The TD output training signals of the PAs are converted to FD symbols constellations. The measured Error Vector Magnitudes (EVMs) of the 16-QAM constellations of the five transmitters are -16.04dB, -17.13dB, -15.94dB, -16.99dB and -16.95dB respectively, which basically satisfy the normal digital communication specifications.

So, the five transmitters can be considered from the same model and same series with subtle differences of aging hardware.

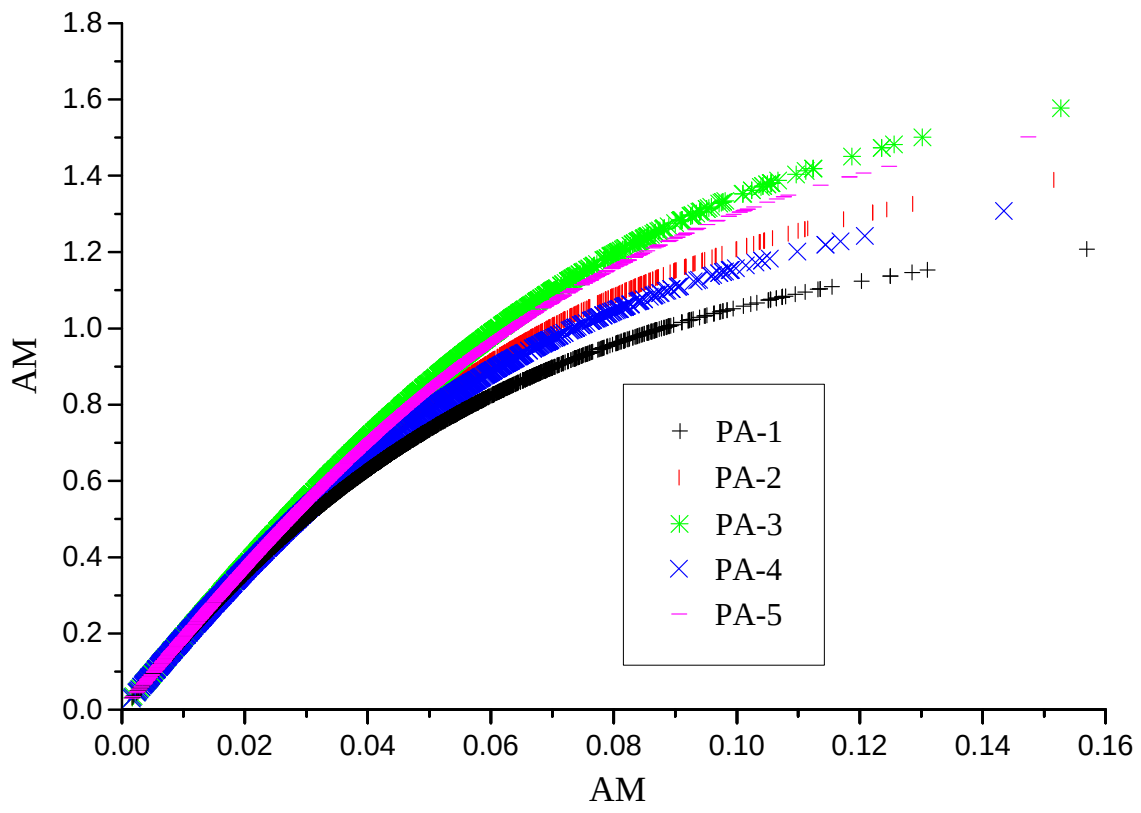

(a)

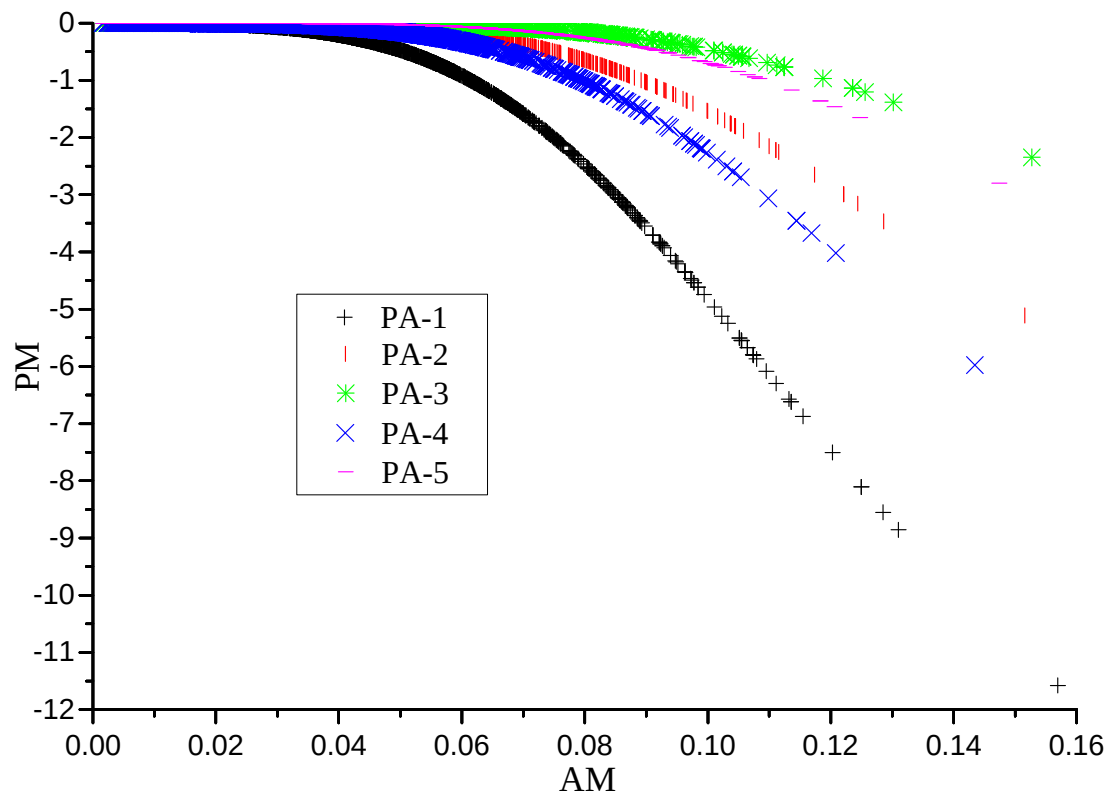

(b)

Figure 3 The nonlinear characteristics of five PAs (a) AM-AM (b) AM-PM

### *B. The classification of the five transmitters*

A basic *k*-NN classifier is used to classify the five transmitters with the extracted feature vectors $\mathbf{F}_1$, $\mathbf{F}_2$ and $\mathbf{F}_{\mathrm{e}}$, where $\mathbf{F}_{\mathrm{e}}$ is for the performance comparison.

In one independent experiment, 66 samples of $\mathbf{F}_1$, $\mathbf{F}_2$ and $\mathbf{F}_{\mathrm{e}}$ are obtained for each transmitter respectively. When $E_b/N_0$=15dB, the feature scatter of $\mathbf{F}_1$, $\mathbf{F}_2$ in one experiment are shown in Fig. 4 (a) and (b), respectively.

It can be seen with Fig. 4 that the five transmitters can be classified easily with $\mathbf{F}_1$ and $\mathbf{F}_2$ respectively. Obtain the ahead 33 samples as the training data, the later 33 samples as the testing data. The correct classification rates are illustrated in Table 2 with the *k*-NN classifier when *k*=1,2,3,4 respectively.

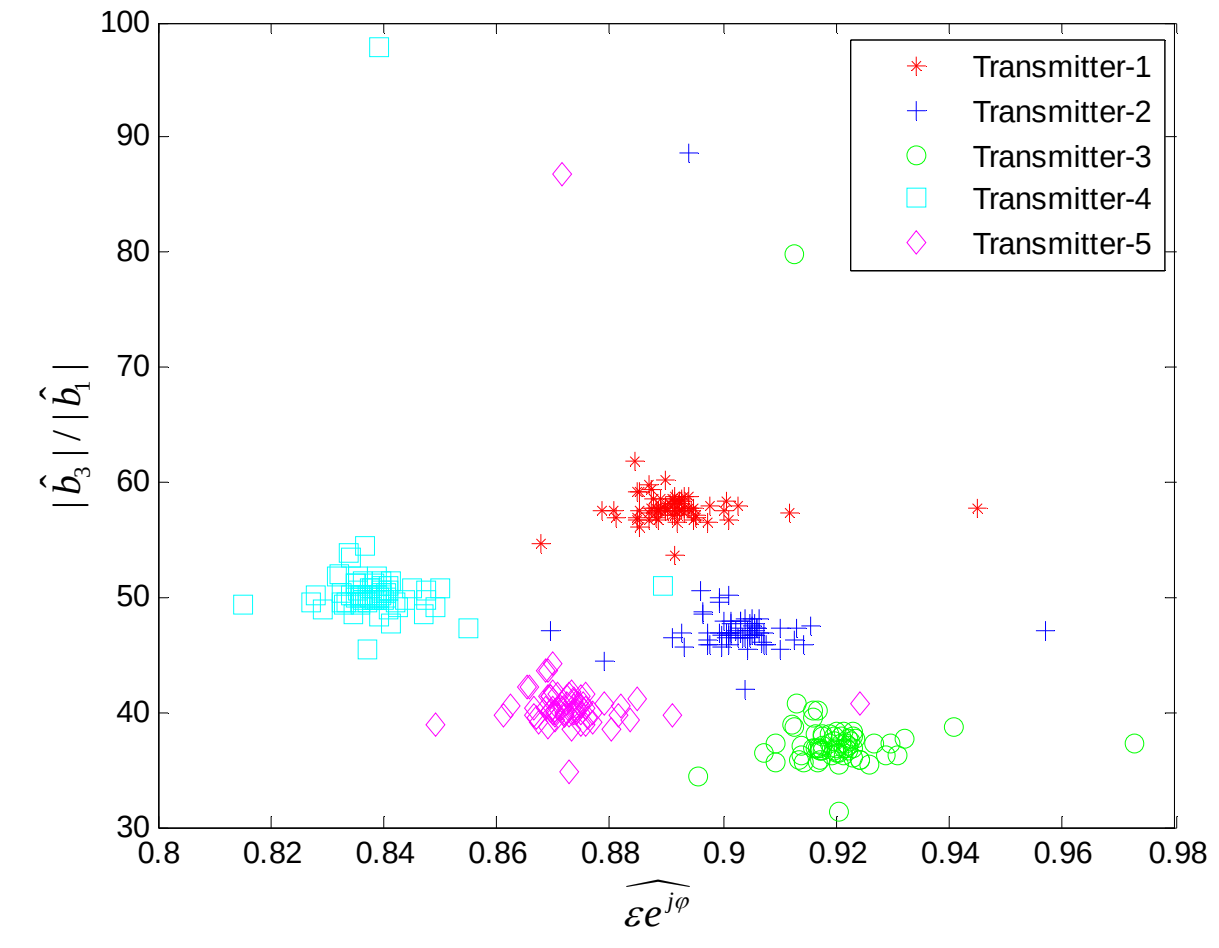

(a)

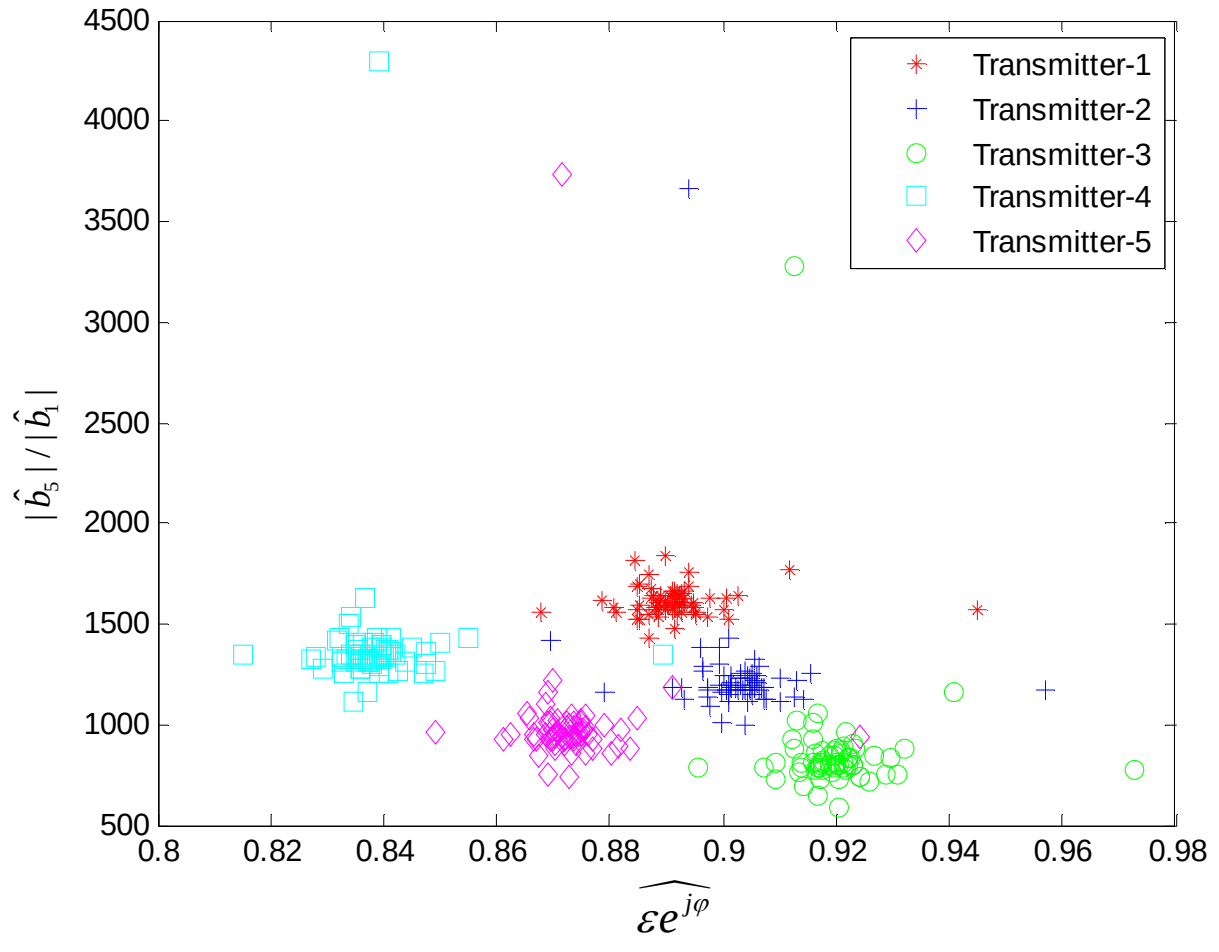

(b)

Figure 4 The scatter of IQ imbalance and nonlinear feature vector in one independent experiment under $E_b/N_0$=15dB (a) $\mathbf{F}_1$ (b) $\mathbf{F}_2$

TABLE II

CORRECT CLASSIFICATION RATES IN ONE INDEPENDENT EXPERIMENT UNDER EB/N0=15DB (%)

| Feature vectors | 1-NN | 2-NN | 3-NN | 4-NN |
|---|---|---|---|---|
| $\mathbf{F}_1$ | 87.27 | 86.06 | 89.70 | 89.09 |
| $\mathbf{F}_2$ | 72.73 | 70.30 | 78.79 | 73.33 |

It can be seen with Table 2 that the correct classification rates with $\mathbf{F}_1$ and $\mathbf{F}_2$ in this independent experiment under $E_b/N_0$=15dB are between 70% to 90% under different $k$ of $k$-NN classifier.

On the other hand, the feature scatters of $\mathbf{F}_e$ in the same experiment are shown in Fig. 5.

It can be seen with Fig. 5 that the five transmitters can not been classified with $\mathbf{F}_e$ at all. Using the same $k$-NN classifier, the correct classification rates with $\mathbf{F}_e$ are 23.64%, 22.42%, 21.21% and 23.64% respectively when $k$ =1 to 4. The reason is that $\mathbf{F}_e$ is directly from the absolute of the received signal which contains the effect of the time variant multipath channel, but $\mathbf{F}_1$ and $\mathbf{F}_2$ are from the estimations of the IQ imbalance and nonlinear model coefficients that are determined by the hardware property of the transmitters. Consequently, the correct classification rates with $\mathbf{F}_e$ are near the random guess probability of the five transmitters, yet that with $\mathbf{F}_1$ and $\mathbf{F}_2$ are greatly higher than the random guess probability in this independent experiment.

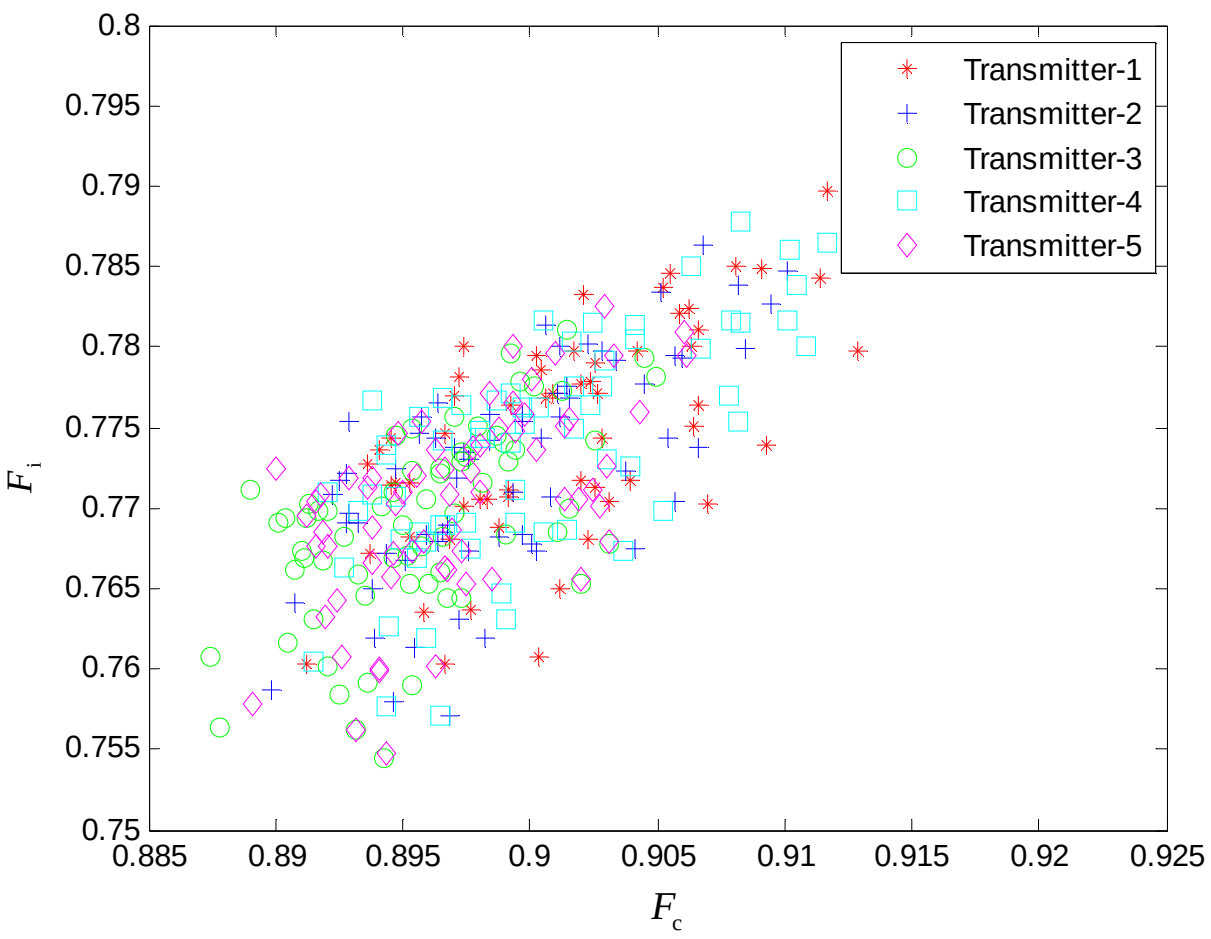

Figure 5 The scatter of envelope feature vector $\mathbf{F}_e$ in the same independent experiment

The 100 MontCarlo experiments with random AWGN and multipath channel are done. $E_b/N_0$ is set from 0dB to 30dB with the step of 5dB. Each experiment under each $E_b/N_0$ generates 66 samples and obtain one group of correct classification rates. The mean of the obtained correct classification rates are depicted in Fig. 6.

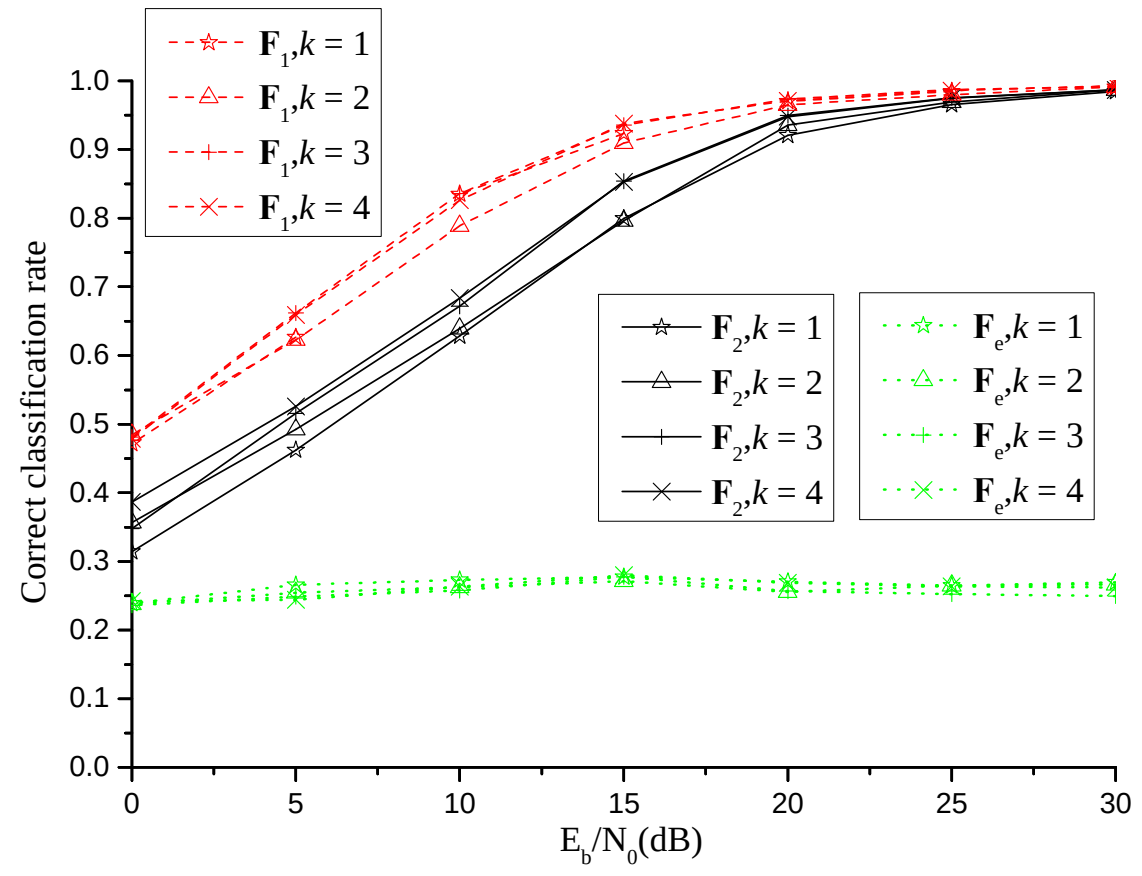

Figure 6 Comparison of the correct classification rates with different feature vectors under different $E_b/N_0$

It can be seen with Fig. 6 that the trend of the correct classification rates of $k$-NN when $k$ = 1, 2, 3, 4 are the same for each feature vector.

It can be seen with Fig. 6 that the correct classification rates with $\mathbf{F}_e$ are the random guess probability or so, 0.25, at all $E_b/N_0$. And It can be seen that the correct classification rates with $\mathbf{F}_1$ and $\mathbf{F}_2$ are all higher than the random guess probability at all $E_b/N_0$. The correct classification rates with the proposed RF Fingerprinting method go up with the increase of $E_b/N_0$. When $E_b/N_0$ goes up to 20dB, the correct classification rates are above 80%. On the other hand, when $E_b/N_0$ drops down than 5dB, the correct classification rates with the novel RFF are below 70%.

## V. Conclusion

In this paper, a novel RF Fingerprinting method for wireless OFDM transmitters based on the hardware property of the nonlinearity and IQ imbalance of the transmitter was proposed. The novel method was limited to the OFDM system using the anti-symmetry training pilots. Yet, the novel RFF is stable as the effect of wireless multipath channels has been eliminated. Even for the OFDM devices from the same model and same series, the novel RF Fingerprinting method is feasible, which has been demonstrated by numerical experiences of five transmitters with minor hardware differences. The proposed method can be used for the security enhancement of the physical-layer authentication in wireless OFDM communications.